\documentclass[aps,twocolumn,prl,showpacs]{revtex4}
\usepackage{graphicx}
\usepackage{amsmath}
\usepackage{epsfig}
\usepackage{dcolumn}
\usepackage{bm}

\begin{document}

\title{Electron Doping of Cuprates via Interfaces with Manganites}

\author{S. Yunoki}
\affiliation{
Department of Physics and Astronomy, The University of Tennessee, 
Knoxville, Tennessee 37996, USA}
\affiliation{
Materials Science and Technology Division, Oak Ridge National Laboratory, 
Oak Ridge, Tennessee 32831, USA.}

\author{S. Okamoto}
\affiliation{
Materials Science and Technology Division, Oak Ridge National Laboratory, 
Oak Ridge, Tennessee 32831, USA.}

\author{S. S. Kancharla}
\affiliation{
Materials Science and Technology Division, Oak Ridge National Laboratory, 
Oak Ridge, Tennessee 32831, USA.}

\author{A. Moreo}
\affiliation{
Department of Physics and Astronomy, The University of Tennessee, 
Knoxville, Tennessee 37996, USA}
\affiliation{
Materials Science and Technology Division, Oak Ridge National Laboratory, 
Oak Ridge, Tennessee 32831, USA.}

\author{E. Dagotto}
\affiliation{
Department of Physics and Astronomy, The University of Tennessee, 
Knoxville, Tennessee 37996, USA}
\affiliation{
Materials Science and Technology Division, Oak Ridge National Laboratory, 
Oak Ridge, Tennessee 32831, USA.}

\author{A. Fujimori}
\affiliation{Department of Physics, University of Tokyo,
7-3-1 Hongo, Bunkyo-ku, Tokyo 113-0033, Japan.}

\date{\today}

\begin{abstract}
The electron doping of undoped high-$T_c$ cuprates via the transfer of charge
from manganites (or other oxides) using heterostructure geometries is here
theoretically
discussed. This possibility is mainly addressed via a detailed analysis
of photoemission and diffusion voltage experiments, which locate the Fermi level
of manganites above the bottom of the upper Hubbard band of some cuprate parent
compounds.  A diagram with the relative location of Fermi levels and gaps
for several oxides is presented. The procedure discussed here is generic, 
allowing for 
the qualitative prediction of the charge flow direction at several oxide interfaces.
The addition of electrons to antiferromagnetic Cu oxides
may lead to a 
superconducting state at the interface with minimal quenched disorder.
Model calculations using static and dynamical mean-field theory, supplemented
by a Poisson equation formalism to address charge redistribution at the interface,
support this view. The magnetic state of the manganites could be antiferromagnetic
or ferromagnetic. The former is better to induce superconductivity than the latter, 
since the spin-polarized charge transfer will 
be detrimental to singlet superconductivity.
It is concluded that in spite 
of the robust Hubbard gaps, the electron doping of undoped cuprates
at interfaces appears possible, and its realization may open an exciting 
area of research in oxide heterostructures.
\end{abstract}


\maketitle

\section{Introduction} 

The study of oxide heterostructures is rapidly developing into one of the most promising areas
of research in strongly correlated electronic systems. The current excitement in  this
field was in part triggered by the recent discovery 
of conducting interfaces, with a substantial high carrier mobility, between two insulating 
perovskites \cite{ohtomo02,ohtomo04}. 
These results were obtained by growing abrupt layers of the insulators LaTi(3+)O$_3$ and SrTi(4+)O$_3$. 
When the spatial distribution of the extra electron was observed with an atomic-scale electron beam, it was  
found to correspond to a metallic state at the interface \cite{ohtomo02}.
Theoretical investigations of these 
systems \cite{okamoto04a,okamoto04b,okamoto2004}, 
using Hartree-Fock and DMFT techniques, concluded that the leakage of charge from one layer to the other explains the results.
Thus, charge transfer between materials in heterostructures can be used to stabilize interface states that otherwise
would only be obtained via the chemical doping of the parent oxide compound. This last procedure has the concomitant 
effect of introducing quenched disorder into the sample. Then, for the preparation of doped materials without the extra complication
of disorder, for the realization of two-dimensional versions of perovskites, 
and for the exploration of ``oxide electronics'' devices and functionalities, 
the experimental and theoretical analysis of oxide heterostructures is a promising
and vast field, and its study is only now starting \cite{nakagawa,thiel,dagotto}.

Before the recent developments mentioned above, 
ferromagnetic/superconducting (FM/SC) heterostructures 
had already received considerable attention.
The analysis of spin injection from 
a ferromagnetic material to an electrode is of much importance for further progress in the area 
of spintronics. Other properties of interest 
in FM/SC heterostructures include: (i) spin-mixing effects 
that can induce spin-triplet pairing correlations at the 
interface \cite{yoshida}; (ii) Josephson couplings between two 
singlet-superconducting layers separated by a half-metallic 
ferromagnet \cite{eschrig};  (iii) oscillations in the singlet pairing 
amplitude detected in SC/FM/SC geometries \cite{kontos}; 
and (iv) magnetic exchange coupling in FM/SC/FM geometries \cite{vasko,melo}.
Since transition metal oxides have similar lattice spacings, the combination 
of colossal magnetoresistance (CMR) manganites \cite{dagotto,dagotto-CMR,book}
and high-$T_c$ cuprates
has been specially investigated \cite{mannhart,ahn,pavlenko}. 
Experimental results suggest a strong FM/SC interplay resulting in 
the injection of 
spin polarized carriers into the SC layers \cite{vasko}. 
Of particular importance for the purposes of
our investigations, the transfer 
of charge from the La$_{1-x}$Ca$_{x}$MnO$_3$ (LCMO) ferromagnet  to the YBa$_2$Cu$_3$O$_y$ (YBCO) 
superconductor has been
observed in recent experiments using EELS techniques \cite{varela}. In this case, the addition of electrons, namely the
suppression of holes, in the YBCO component led to the concomitant suppression of superconductivity and a transition to
an insulating state. Similar effects
are also believed to occur in grain boundaries of superconductors \cite{hilgenkamp}. In other
LCMO/YBCO superlattice setups,
the superconducting critical temperature was also found to be suppressed \cite{sefrioui}.
This may be caused by the influence of spin-polarized
carriers moving from LCMO to YBCO, 
breaking singlet Cooper pairs, or by the reduction of the number of hole carriers \cite{varela},
or both.

It is the main purpose of the investigation reported here
to analyze whether the previously observed electronic charge transfer at interfaces from  
manganites, such as LCMO or La$_{1-x}$Sr$_{x}$MnO$_3$
(LSMO), to hole-doped cuprates, such as YBCO, 
can be extended to the case where the cuprate is in an undoped state. If this still occurs, 
namely if estimations of work
functions lead us to believe that the Fermi level of the manganite remains above the first empty states of 
the undoped cuprate, then this
Cu oxide will receive an excess of electrons, leading to the $electron$ $doping$ of an antiferromagnetic
(AF) state at heterostructure interfaces \cite{oh}. This
 mixing of cuprates and manganites could lead either
to (i) a metallic spin-polarized state at  the interface, if the manganite is ferromagnetic, 
or to  (ii) an electron-doped superconductor at low
enough temperatures, if the spin polarization is not strong enough to destroy Cooper pairs or if the manganite
used is not ferromagnetic. 
This mechanism would allow for the electronic doping of a parent cuprate compound
without the complication
of the Coulombic and structural disorder introduced by chemical doping. The fact 
that this is electron doping, as opposed to the already investigated interfacial hole doping, is particularly
interesting since a variety of material issues has prevented the electron-doped branch 
of high-$T_c$ superconductors to develop as vastly as those
that are hole doped. 

Once again, note that if the manganite is ferromagnetic, then the charge transfer to the
undoped cuprate would occur via polarized electrons, which
suppress spin singlet 
superconductivity. Thus, there are two $competing$ tendencies: adding electrons 
favors superconductivity, but its
spin polarization suppresses it. The outcome is difficult to predict. Recent investigations
using realistic microscopic models focusing on a 
manganite/insulator interface predict that a complex pattern of phase separation between FM and AF
regions could occur \cite{brey}.
Thus, even if superconductivity is not induced, the resulting state would be interesting,
due to the interplay between ferro and antiferromagnetic tendencies.
More importantly, note that 
the manganite involved in the heterostructure does $not$ need to be ferromagnetic: low-bandwidth Mn oxides
present a wide variety of spin/charge/orbital order 
states, other than ferromagnetism \cite{dagotto-CMR,book}. Even large-bandwidth manganites have 
non-FM states at small and large enough hole concentration. Thus, inducing a SC
state is a real possibility
with the proper choice of the manganite partner to the cuprate.

\begin{figure}[hbt]
\includegraphics[clip=true,width=7.5cm,angle=-0]{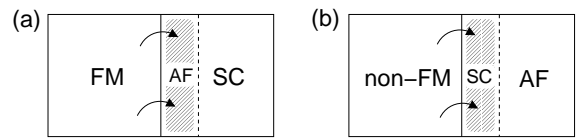}
\begin{center}
\caption{
Schematic representation of the systems studied here.
(a) is a heterostructure involving a ferromagnetic 
(FM) half-metal system, such as LCMO, and a superconductor (SC), such as YBCO. In this case, an antiferromagnetic 
(AF) insulating state is induced at the interface. (b) is a heterostructure involving a non-FM manganite 
and an undoped AF cuprate. Here, the transfer of charge may lead to a superconducting state at the
interface. Arrows indicate the direction of flow of electrons, which are accumulated 
in the gray region. 
}
\label{systems}
\end{center}
\end{figure}

A qualitative representation of the transfer of
charge is shown in Fig.~\ref{systems}
 for the cases of a doped cuprate, such as discussed in \cite{varela}, and for an undoped cuprate,
the focus of our investigation reported below. In case (a), SC is destroyed, while in the proposed case (b), SC is induced (if the
manganite is not ferromagnetic).
Our main result is that the novel scenario (b) appears possible. Our conclusion is based 
on (i) a detailed analysis of the work functions of the materials
of relevance for these investigations, mainly involving the study of  photoemission and diffusion voltage published data, as well
as (ii) rough
calculations at the mean-field and dynamical mean-field levels, showing 
that there is no fundamental problem preventing superconductivity from happening at the
interface. However, it must be clearly expressed that our contribution should be considered as only the
first theoretical steps toward realizing electron doping at cuprate interfaces,
and considerable
more  work remains to be carried out. For instance, issues such as lattice 
reconstructions and polarity of the involved interfaces  have $not$ been analyzed here (only electronic reconstructions
at idealized perfect interfaces were investigated). Vacancies and the most optimal location of oxygens at the interface
have not been investigated as well.
However, even with these caveats we believe the conclusions described below are sufficiently
interesting that they deserve experimental efforts to attempt their realization in real heterostructures.

The organization of the manuscript is as follows: first, we carry out a detailed analysis
of available experimental information to judge if the Fermi level of manganites is above
the lowest-energy empty state of several undoped cuprates. This is described in detail, because
it provides a systematic procedure to address other oxide heterostructures in the future.
Second, a model calculation in the mean-field approximation is carried out, both for 
static and dynamical cases. No fundamental problem is found with the proposal of having
superconductivity induced at the manganite/cuprate interface. The paper concludes
with a discussion and summary.

\section{Estimation of Oxides Chemical Potential Differences
using Photoemission and Diffusion Voltage Experiments}

\subsection{Overview}

For the success of the emergent field of oxide heterostructures it is crucial to properly
determine the relative work functions of the materials involved, since these work functions control
the curvature of the valence and conduction bands (VB and CB) of the constituent materials, 
and ultimately the carrier concentration at the interfaces. 
Work functions of conventional metals and semiconductors have been studied for decades 
establishing the fundamental background for current electronics. 
Thus, for the next-generation 
electronic devices utilizing the complex properties of correlated-electron systems,
such as high-$T_c$ cuprates and CMR manganites, determining the
relative work functions of a variety of transition-metal oxides is equally important.

Photoemission spectroscopy (PES) is an important technique in this context. 
PES has provided fundamental
information to uncover the properties of complex oxides. 
However, although PES techniques can be of considerable help for oxide heterostructures, 
only a limited number of experiments have been reported specifically addressing
the work function of transition-metal oxides \cite{Schulte01}.

By measuring the diffusion voltage $V_d$ (or built-in potential) of a junction between two materials of
interest,  one can also extract the chemical potential differences between them. 
This diffusion voltage is the potential barrier at the interface 
after the rearrangement of charge occurs.
If there are no extra contributions to $V_d$, such as those caused by
interface polarities, impurities, or lattice reconstructions, then 
$V_d$ is equivalent to the work function difference between the two constituents. 
If the work function of one of the materials is known, the work function of the other can be estimated. 
This provides another procedure to study the band alignment of oxides. 

The purpose of this section is to provide a rough estimation of the band diagram of 
perovskite transition-metal oxides (cuprates and manganites) using
the experimental data currently available. This is achieved by
combining information from chemical potential shifts obtained
using PES with diffusion voltage measurements on heterostructures. 

\subsection{Parent compounds of High-$T_c$ cuprates}

Since the discovery of high-$T_c$ superconductivity in the cuprates, 
considerable experimental information has been accumulated about these compounds.
In particular, the chemical-potential jump from the
hole-doped to the electron-doped materials has been one of the important topics since
it is closely related to the Mott/Hubbard (or charge transfer) gap.
Here, let us first consider La$_2$CuO$_4$ (LCO) and Nd$_2$CuO$_4$ (NCO)
as the parent compounds of hole-doped 
and electron-doped cuprates, respectively. 
It should be mentioned that these compounds have different structures:
LCO has the layered perovskite or $n=1$ Ruddlesden-Popper structure, 
while NCO presents the so-called $T'$-structure without apical oxygen. 
Even though the electronic properties of the 
CuO$_2$ planes are quite similar for LCO and NCO,  the different overall structures
suggest that small differences may exist in their chemical potentials, as discussed below.

From the optical measurements, it has been suggested  that the chemical potential of 
LCO is located 0.4~eV above the top of the valence band \cite{Suzuki89}. 
Similar results were obtained
with PES techniques \cite{Ino00}. Polaronic effects are believed to be the cause of this
somewhat ``mysterious'' shift between the top of the valence band and the actual location
of the chemical potential in LCO \cite{polarons}.
Similarly,
resonant-photoemission studies on NCO revealed that the chemical potential of NCO 
is located 0.7~eV above the valence band \cite{Allen90,Namatame90}. 
If the positions of the valence bands are identical between LCO and NCO, 
then their chemical potential jump becomes $0.7-0.4=0.3$~eV  \cite{alternative}.  
Although an accurate determination of the individual chemical 
potentials of these materials remains to be done, 
estimations based on the shifts of the O 1$s$ and Cu 2$p$ core levels indicate that
the chemical potential jump is at most 0.5~eV \cite{Fujimori02}. Then,
the positions of the valence bands can be assumed to be the same, although with a  $\alt 0.2$~eV
uncertainty.

\begin{figure}[tbp]
\includegraphics[width=0.55\columnwidth,clip]{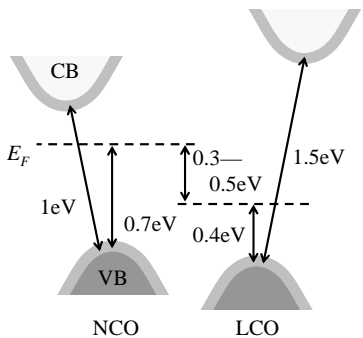}
\caption{Schematic band diagrams of NCO and LCO \cite{Fujimori02} 
based on the chemical potential jump between NCO and LCO found in a resonant-photoemission study \cite{Namatame90},
revised by considering the recent discovery of an indirect Mott gap in the cuprates \cite{Hasan00}. 
Chemical potentials (dashed lines) of NCO and LCO are located about 0.7~eV and 0.4~eV above the 
valence bands, respectively. 
Note that the top of the valence bands of NCO and LCO do not necessarily match, but estimations
discussed in the text locate them very close to one another.
There remains an ambiguity in the chemical potential jump between these two materials, 
but it is believed to be at most 0.5~eV \cite{Fujimori02}. For more details see text.
}
\label{fig:NCO_LCO}
\end{figure}

To complete the band diagram, information about the location of 
the unoccupied conduction, or upper Hubbard band, is needed. 
From optical spectroscopy, it was first suggested that the 
Mott gaps of LCO and NCO are about 2~eV and 1.5~eV, respectively \cite{Tajima89}.
However, numerical studies of the
Hubbard model on finite-size clusters using  appropriate hopping amplitudes 
for the cuprates revealed that the Mott gap in Cu oxides 
was in fact an indirect one \cite{Tsutsui99}: 
the top of the lower-Hubbard band is at momenta
$(\pm \pi/2,\pm \pi/2)$ while the bottom of the upper-Hubbard band lies at $(\pi,0)$ and $(0,\pi)$.
Thus, the Mott gap is indirect and smaller than the optical gap. 
The indirect nature of this gap has been confirmed by a recent resonant x-ray study \cite{Hasan00}.
Considering the ratio between the magnitudes of direct gap and indirect gap from the 
theoretical considerations and the optical gap obtained by the experiments, 
the separation between the VB and CB of LCO and NCO is estimated to be about 1.5~eV and 1~eV, respectively \cite{BIS98}. 
Combining all this information together, 
the schematic band diagram for these materials can be constructed and it is shown in Fig.~\ref{fig:NCO_LCO}.

Next, let us address another series of high-$T_c$ cuprates:  YBCO. 
The diffusion voltage of the heterostructure between YBCO (oxygen contents unclear) 
and 0.05wt\% Nb-doped SrTiO$_3$ (STO) (Nb$_{0.05}$-STO)
has been measured \cite{Muraoka04}. 
The chemical potential of YBCO was found to be 1.5~eV below that of Nb$_{0.05}$-STO. 

To establish
a connection with the single-layer cuprates, note that
similar experiments on the diffusion voltage of the single-layer parent compound Sm$_2$CuO$_4$ (SCO) and 
0.01wt\% Nb-doped SrTiO$_3$ (Nb$_{0.01}$-STO) have been very recently performed \cite{Nakamura07}. 
SCO is an isostructural material of NCO, with the same $T'$-structure and formal valences, therefore 
NCO is expected to have a chemical potential very similar to that of SCO. 
The recent experiments of Nakamura {\it et al.}
revealed that the chemical potential of SCO is 1.3~eV below that of Nb$_{0.01}$-STO \cite{Nakamura07}. 
Although there is a slight difference in the doping
concentration of Nb used for the two experiments (0.05wt\% Nb for YBCO  and 0.01wt\% for SCO), 
by reducing the Nb concentration to 0.01wt\% the diffusion voltage is expected to increase 
only by 0.1~eV or less because of band-gap narrowing effects \cite{SawaPC}. 
Thus, by combining the diffusion voltages on SCO- and YBCO-based heterostructures 
involving a Nb-doped STO substrate, 
the chemical potential of SCO is estimated to be $\sim 1.5+0.1-1.3 = 0.3$~eV above that of YBCO. 

\begin{figure*}[tbp]
\includegraphics[width=1.5\columnwidth,clip]{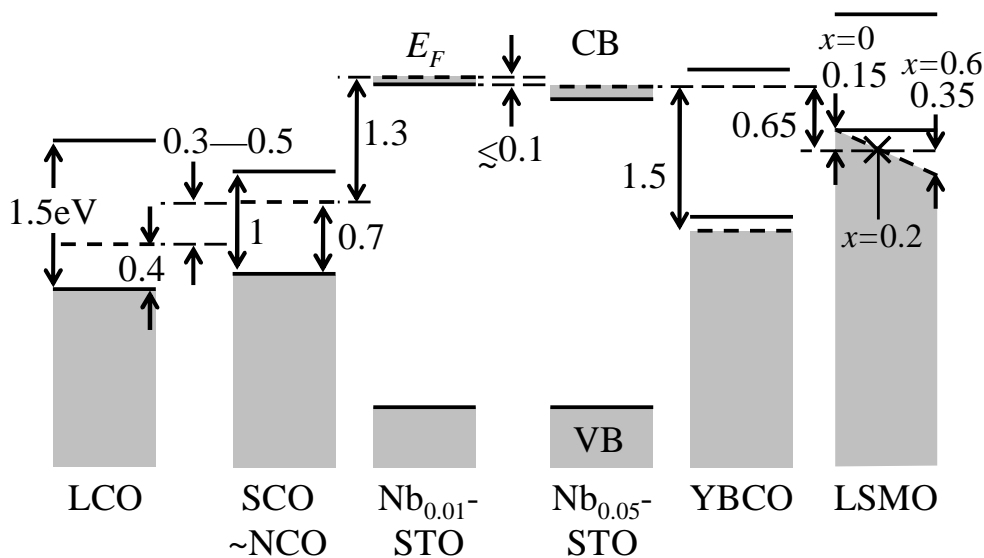}
\caption{Schematic band diagrams of LCO, SCO(NCO), Nb$_{0.01}$-STO, Nb$_{0.05}$-STO, YBCO, and LSMO 
based on diffusion voltage measurements \cite{Muraoka04,Muramatsu05,Nakamura07,SawaPC} 
and photoemission spectroscopy \cite{Namatame90,Fujimori02,Matsuno02}.
Tops of valence bands (VB) and bottoms of conduction bands (CB) are indicated by solid lines, 
while chemical potentials are indicated by dashed lines.}
\label{fig:LCO_SCO_NbSTO_YBCO_LSMO}
\end{figure*}

Furthermore, assuming that the behavior of SCO and NCO are identical, based
on the similarity of their structures, then the
chemical potential of LCO can be estimated by considering the chemical potential jump between NCO and LCO. 
Figure~\ref{fig:LCO_SCO_NbSTO_YBCO_LSMO} summarizes 
the band diagram for the various high-$T_c$ cuprates studied here.
From these results, 
it is deduced that the parent compounds of hole-doped cuprates have fairly similar chemical potentials 
despite their different crystal structures. 

\subsection{ Doping dependence of the chemical potential in High-$T_c$ cuprates}
\begin{figure*}[tbp]
\vskip 1.0cm
\includegraphics[width=1.75\columnwidth,clip]{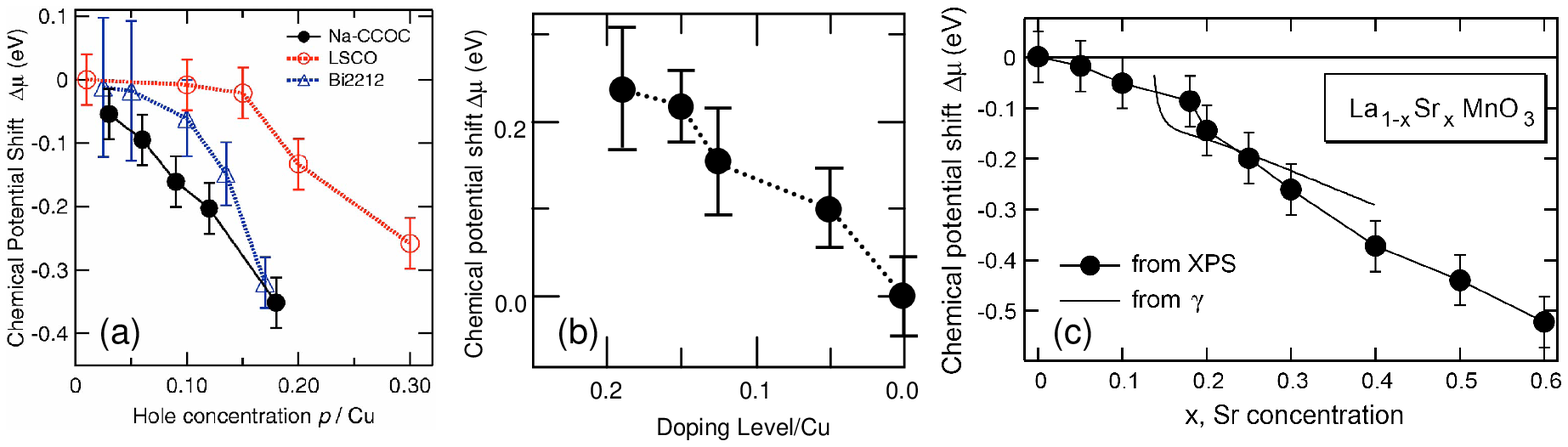}
\caption{Chemical-potential shifts of various transition-metal oxides, reproduced
from photoemission experiments. 
(a) are the results for hole-doped cuprates. 
Na-CCOC: Ca$_{2-x}$Na$_x$CuO$_2$Cl$_2$ \cite{Yagi06}, 
LSCO: La$_{2-x}$Sr$_x$CuO$_4$ \cite{Ino97},
and Bi2212: Bi$_2$Sr$_2$Ca$_{1-x}$R$_x$Cu$_2$O$_{8+y}$ (R=Pr, Er) \cite{Harima03}.
The absolute values at the undoped origin of each series are assumed to be the same.
Reproduced from \cite{Yagi06}. 
(b) Chemical potential shift of electron-doped cuprates NCCO: 
Nd$_{2-x}$Ce$_x$Cu$_2$O$_4$.
Reproduced from \cite{Harima01}. 
(c) Chemical-potential shift of LSMO.  
These shifts are measured at liquid-nitrogen temperature except for LaMnO$_3$. Reproduced
from \cite{Matsuno02}. 
}
\label{fig:WorkFunctions}
\end{figure*}

The doping dependence of the chemical potential is important when using doped compounds for the
oxide heterostructures. 
The chemical potential shifts from their parent compounds 
of various high-$T_c$ cuprates  have 
been intensively studied \cite{Namatame90,Ino97,Harima03,Yagi06}. 
For the benefit of the readers,
the main results are summarized
in Fig.~\ref{fig:WorkFunctions} (a) for hole-doped compounds, 
and Fig.~\ref{fig:WorkFunctions} (b) for electron-doped compounds \cite{Harima01}.

There are still considerable discussions 
on the electronic properties of the underdoped region of La$_{2-x}$Sr$_x$CuO$_4$ (LSCO), involving concepts such as 
stripe formation, phase separation, and spin glass. 
Those may explain the flat chemical potential shifts of LSCO at $0\le x < 0.15$. 
However, focusing on the overdoped region where the mixed-phase complexity is reduced, then the 
chemical potential shifts behave linearly with hole doping
with a slope of about 0.2 eV per 0.1 carrier concentration, which is close to that of 
electron-doped Nd$_{2-x}$Ce$_x$CuO$_4$ (NCCO). 
This may indicate that the band structures of the CuO$_2$ planes in those compounds are similar, as widely believed.
On the other hand, the
 $\Delta \mu$ vs. carrier concentration slope for Bi$_2$Sr$_2$CaCu$_2$O$_{8+y}$ (Bi2212) is larger than
for the single-layer compounds, perhaps due to its bilayered nature.

\subsection{ Connecting  the High-$T_c$ cuprates with the CMR manganites}
As already discussed, both from the fundamental physics as well as the device engineering perspective, 
heterostructures involving high-$T_c$ cuprates and CMR manganites are important.
Here, we consider the cubic perovskite manganites first, and then turn to the double-layered manganites. 

It should be mentioned that the $direct$ measurement of the work function 
of cubic manganites has been performed by several groups 
using PES \cite{Jong03,Minohara07} and the Kelvin method \cite{Reagor04}, 
and results are approximately consistent with one another indicating an intrinsic work function 
$\sim 4.8$~eV for La$_{1-x}$Sr$_x$MnO$_3$ with $x=0.3,0.4$. 
These experimental data are important in considering interfaces with the cuprates. 
However, as the individual 
work functions of many cuprates are not yet available but only their differences, then we 
have to consult other experiments in order to establish a connection between 
manganites and cuprates. 
Here, we consider PES experiments that measured the chemical potential shift in doped manganites \cite{Matsuno02}, 
and those that provided the diffusion-voltage  
on La$_{0.8}$Sr$_{0.2}$MnO$_3$ (20\% hole-doped LSMO)/Nb$_{0.05}$-STO heterostructures \cite{Muramatsu05}. 

According to the experiment by Matsuno {\it et al}. \cite{Matsuno02}, the
chemical potential of La$_{1-x}$Sr$_x$MnO$_3$ (LSMO) changes rather monotonically from the parent compound LaMnO$_3$ to 
a value $\Delta \mu$=$-0.5$~eV for La$_{0.4}$Sr$_{0.6}$MnO$_3$ (60\% hole-doped LSMO)
(see Fig.~\ref{fig:WorkFunctions} (c)).
Furthermore, diffusion voltage measurements 
on a La$_{0.8}$Sr$_{0.2}$MnO$_3$ (20\% hole-doped LSMO)/Nb$_{0.05}$-STO heterostructure,
done by Muramatsu {\it et al.} \cite{Muramatsu05}, reported a value $V_d = 0.65$~eV. 
Combining these experiments, the
bands of LSMO can be aligned relative to those of the cuprates obtained in the previous section. 
Figure \ref{fig:LCO_SCO_NbSTO_YBCO_LSMO} summarizes the band alignment of cuprates and manganites, 
which is our main result of this section. 
In particular, the chemical potential shifts between the various cuprates of potential
relevance and the 20\% doped LSMO are summarized as follows:  
(i) for SCO $\Delta \mu = \mu_{LSMO} - \mu_{SCO}\sim 0.55$~eV,
(ii) for LCO $\Delta \mu = \mu_{LSMO} - \mu_{LCO}\sim 0.85$--$1.05$~eV, 
and (iii) for YBCO $\Delta \mu = \mu_{LSMO} - \mu_{YBCO}\sim 0.85$~eV. 
In particular, it is natural to expect that electrons be
transfered  from LSMO to the valence band of hole-doped YBCO at an interface between the two materials.
This has already been observed experimentally \cite{varela}, thus it is reassuring 
that the simple procedure followed
here is consistent with available experimental information \cite{YBCO-work}.

Consider now another class of Mn oxides, the 
double-layered manganites with the so-called $n=2$ Ruddlesden-Popper structure. 
Direct measurements of work function using photoemission have actually been performed on 
La$_{1.2}$Sr$_{1.8}$Mn$_2$O$_7$. 
It has been reported that the work function of La$_{1.2}$Sr$_{1.8}$Mn$_2$O$_7$ 
increases with decreasing temperature across the Curie temperature (125~K) 
from $\sim$~3.5~eV at 180~K to $\sim$~3.56~eV at 60~K \cite{Schulte01}, 
while the simple double-exchange model predicts the opposite trend. 
This fact may indicate the importance of several ingredients neglected in the double-exchange model, 
such as electron-electron interaction, electron-lattice interaction, and orbital degeneracy. 
The important point worth emphasizing here is that 
the work function of the double-layered manganite 
is more than 1~eV smaller than that of cubic LSMO 
(whose work function is $\sim4.8$~eV) \cite{Jong03,Minohara07,Reagor04}, 
which means that the chemical potential of the 
double-layer manganite is more than 1~eV $higher$ than the cubic perovskite LSMO. 
Thus, at least naively,
it should be possible to inject electrons to the undoped cuprates using the double-layer manganites. 
However, there are issues that remain to be clarified for this conclusion to be valid, particularly
the role of the surface condition:
for double-layered manganites, the cleaved surface used in PES experiments 
is always AO$^{2-}$ (A: A-site ion, such as La$^{3+}$ and Sr$^{2+}$),
thus positively charged. 
The low work function of
bilayered manganites could be caused by this charged surface.
On the other hand, for cubic manganites the surface can be either positively charged (AO termination)
or negatively charged (MnO$_2$ termination). 
For this purpose, photoemission experiments on cubic manganites with controlled surface conditions 
are highly desirable. 

Summarizing, in this subsection we discussed the possible electron doping of cuprates from manganites at  ideal interfaces. 
In the paragraphs above, it was concluded that 
the parent compounds of electron-doped cuprates, such as NCO and SCO, are the best candidates for this purpose
since they have the lowest conduction band
(the CB in LCO is about 0.5~eV higher than that in NCO, 
and the CB in YBCO is expected to be even higher than those of NCO and LCO). 
To dope electrons into a cuprate parent compound, 
the minimal condition is that the chemical potential of manganites be higher 
than the bottom of the conduction band (or the chemical potential
if they are not the same) of the undoped cuprate involved. 
This condition seems to be satisfied for cubic manganites LSMO 
in a substantial doping range $x$ that includes FM and AF states, when mixed with NCO or SCO \cite{STO-comment}.
Possible improvements may be achieved by using double-layered manganites, 
which have about 1~eV higher chemical potential than cubic manganites.

\section{Superconductivity at the Manganite/Cuprate Interface: Static Mean-Field Approximation}

The simple ideas described in the previous sections on
the possibility of electron doping of cuprates at interfaces
need to be checked in more theoretical detail to confirm their
consistency. For this purpose, first a static mean-field study will
be here discussed, followed in the next section by a more detailed
analysis including dynamical effects. The emphasis will be given
to a manganite/cuprate interface, but the generation of superconductivity
by electron doping can occur for any other combination where
electrons are donated to the cuprates.

The model used here is defined by the following
Hamiltonian on a three-dimensional cubic lattice:
\begin{equation}\label{model}
H=-\sum_{{\bf i},{\bf j}}\sum_s t_{\bf ij}
c_{{\bf i}s}^\dag c_{{\bf j}s} + H_I 
+\sum_{\bf i}\left(\phi_{\bf i}-\mu+W_{\rm L/R} \right)n_{\bf i}. 
\end{equation} 
Here $c_{{\bf i}s}^\dag$ is the electron creation operator at site 
${\bf i}=(i_x,i_y,i_z)$ with spin $s=\uparrow,\downarrow$, $n_{\bf i}$ 
is the electron density at site {\bf i}, and 
$t_{\bf ij}$ is the nearest-neighbor hopping, which is $t$ on the 
$xy$ plane and $t_z$ in the $z$ direction. $\phi_{\bf i}$ is the electronic 
potential (discussed in more detail later) that will take into account effects
related to the charge redistribution. $\mu$ is the chemical potential, and $W_{\rm L}$ 
($W_{\rm R}$) are site potentials 
for the left (right) side of the system to regulate the transfer of charge. 
$H_I$ contains the interaction 
terms. On the left side of the system (the manganite) the interaction is
\begin{equation}\label{dx}
H_I^{\rm (L)} = -J_{\rm H}\sum_{\bf i}\sum_{\alpha,\beta}
c_{{\bf i}\alpha}^{\dag}\left(\vec{\sigma}\right)_{\alpha\beta}
 c_{{\bf i}\beta}\cdot {\vec{S}}_{\bf i}, 
\end{equation}
which is the standard ``double-exchange'' term.
$\vec{\sigma}=(\sigma_x,\sigma_y,\sigma_z)$ are Pauli matrices, and
${\vec{S}}_{\bf i}$ is a classical localized spin at site {\bf i} 
($|{\vec{S}}_{\bf i}|=1$) representing the $t_{\rm 2g}$ spins.
On the right side of the system, the cuprate, 
the standard repulsive Hubbard interaction ($U$$>$0) is 
supplemented by
a nearest-neighbors attraction ($V$$<$0) that favors superconductivity
in the $d$-wave channel:
\begin{equation}\label{uv}
H_I^{\rm (R)} = U\sum_{\bf i}n_{{\bf i}\uparrow}n_{{\bf i}\downarrow} 
+ V \sum_{\langle{\bf i}{\bf j}\rangle}n_{\bf i}n_{\bf j}.
\end{equation}
The number operator is
$n_{{\bf i}s}=c_{{\bf i}s}^{\dag}c_{{\bf i}s}$, 
and $\langle{\bf i}{\bf j}\rangle$ indicates 
a pair of nearest-neighbor sites {\bf i} and {\bf j} on the $xy$ plane. 
It is well known that this $t$-$U$-$V$ model leads to a rich phase diagram,
that includes superconductivity \cite{tuv}.
Periodic (open) boundary conditions are used in the 
$x$ and $y$ ($z$) directions 
for a lattice of size $L=L_x\times L_y\times L_z$ sites. The focus of the
results described below is on $L_x$=$L_y$=16, but other sizes have been
checked confirming that size effects are not strong in this study.

Two comments are in order: (1) A fully realistic model for manganites
should contain two $e_g$-orbitals and
a Jahn-Teller electron-phonon coupling. However, it is common practice in this context 
to use just one
$e_g$-orbital for simplicity, since several phenomena
 are common to both one and two
orbitals \cite{dagotto-CMR}. Moreover, the focus here is not 
on the CMR regime but only on
combining a standard homogeneous manganite state with a cuprate. 
Thus, the electron-phonon (e-ph) interaction, crucial to understand the CMR
effect, is here neglected for simplicity. In fact, LSMO does not have a large
CMR effect, thus the e-ph coupling may not be strong in this material.
(2) More severe is the approximation carried out on the cuprate side
since a mean-field approximation with an explicitly attractive force is employed.
However, this approximation will be relaxed below and the Hubbard
model will be used without an explicit attraction, 
at the price of having to constraint the study to smaller
systems than those that can be analyzed in the static mean-field approximation.

To treat the many-body interactions described by $H_I^{\rm (R)}$, here a simple mean-field 
approximation is adopted. For the first interaction term, the following replacement is introduced,
\begin{equation}
U\sum_{\bf i}n_{{\bf i}\uparrow}n_{{\bf i}\downarrow} \to 
U\sum_{\bf i} \left [
\langle n_{{\bf i}\uparrow}\rangle n_{{\bf i}\downarrow} 
+ n_{{\bf i}\uparrow} \langle n_{{\bf i}\downarrow}\rangle  
- \langle n_{{\bf i}\uparrow}\rangle \langle n_{{\bf i}\downarrow}\rangle 
\right]. \nonumber
\end{equation}
For the second term, a standard BCS model approximation is used:
\begin{equation}
\begin{array}{l}
\displaystyle{
V \sum_{\langle{\bf i}{\bf j}\rangle}n_{\bf i}n_{\bf j} \to}\\
\displaystyle{
V \sum_{\langle{\bf i}{\bf j}\rangle}\left[\left(
\Delta_{\bf ij}c_{{\bf i}\uparrow}^{\dag}c_{{\bf j}\downarrow}^{\dag} 
+ \Delta_{\bf ji}c_{{\bf j}\uparrow}^{\dag}c_{{\bf i}\downarrow}^{\dag}
\right) + {\rm H.c.} +|\Delta_{\bf ij}|^2 + |\Delta_{\bf ji}|^2
\right],\nonumber}
\end{array}
\end{equation}
where 
$\Delta_{\bf ij}=\langle c_{{\bf j}\downarrow} c_{{\bf j}\uparrow}\rangle$. 
The effect of $V$ is restricted to generate in-plane superconducting 
correlations. To further simplify the analysis and to reduce 
computational efforts, two sublattices are assumed in the $xy$ plane since it is
known that this model has a tendency toward antiferromagnetism at half-filling: e.g.,
$\langle n_{{\bf i}\uparrow}\rangle=\langle n_{i_z\uparrow}^{(\rm A)}\rangle$ 
and $\langle n_{i_z\uparrow}^{(\rm B)}\rangle$ for $(-1)^{i_x+i_y}=1$ 
(A-sublattice) and $-1$ (B-sublattice), respectively. More details can be
found in \cite{alvarez}.

To properly
describe a stable junction made out of materials with
different electronic densities,
it is crucial to include long-range Coulomb interactions between electrons 
and also with 
the background of positive charges.
In this paper, the long-range Coulomb 
interactions are treated within the Hartree approximation by using the 
following Poisson's equation~\cite{datta}: 
\begin{equation}\label{poisson}
\nabla^2\phi_{\bf i} 
= -\alpha\left[\langle n_{\bf i}\rangle-n_+({\bf i})\right], 
\end{equation}
where $\alpha=e/\varepsilon a$ ($\varepsilon$, $e$, and $a$ are the 
dielectric constant, electronic charge, and lattice spacing, respectively). 
$n_+({\bf i})$ is the background positive charge density at site {\bf i}. 
In the following, $e$ and $a$ are set to be 1. 
Furthermore, $n_+({\bf i})$ is kept constant to a number
$n_+^{\rm L}$ ($n_+^{\rm R}$) 
on the left (right) side of the system. The Poisson's 
equation (\ref{poisson}) is solved numerically using symmetric discretizations
in the $x$ and $y$ directions, and a 
forward discretization in the $z$ direction, e.g., 
$d^2\phi_{\bf i}/dz^2
=\phi_{{\bf i}+2{\bf z}}-2\phi_{{\bf i}+{\bf z}}+\phi_{\bf i}$ ({\bf z}: unit 
vector in $z$ direction). The use of the Hartree approximation here is not
a severe problem: the Poisson equation is widely considered to be a reasonable
starting point to account for charge transfer, and it is much used in the
study of semiconducting systems.

To mimic a ferromagnetic half-metallic ground state on the left side of the system, 
the localized spins $\vec{S}_{\bf i}$ are chosen to be 
ferromagnetically aligned, pointing 
into the $z$ direction. To mimic an A-type AF state, these classical spins are
chosen with the same orientation in the $xy$ planes but opposite between adjacent planes. 
On the cuprate side, the mean-field Hamiltonian $H_{\rm MF}$ is first 
Fourier transformed 
to momentum space in the $xy$ plane, and the resulting Bogoliubov-De Gennes 
equation~\cite{bdg} is solved numerically 
by diagonalizing a $(4L_z\times4L_z)$ Hamiltonian matrix for each momentum $(k_x,k_y)$. 
The $d$-wave order parameter is
$\Delta_d({\bf i})=\delta({\bf i},{\bf x}) + \delta({\bf i},-{\bf x}) 
- \delta({\bf i},{\bf y}) - \delta({\bf i},-{\bf y})$, where 
$\delta({\bf i},{\bf j}) = 
\langle c_{{\bf i}\uparrow}c_{{\bf i}+{\bf j}\downarrow} - 
c_{{\bf i}\downarrow}c_{{\bf i}+{\bf j}\uparrow} \rangle /2$, and 
{\bf x} and {\bf y} are unit vectors in the 
$x$ and $y$ directions, respectively.

\subsection{FM-SC interface: charge transfer leading to AF}
\begin{figure}[hbt]
\includegraphics[clip=true,width=7.5cm,angle=-0]{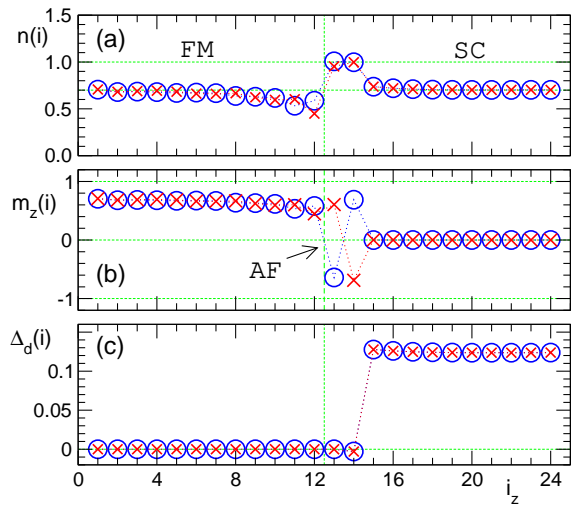}
\begin{center}
\caption{
(Color online) Transfer of charge from a FM half-metal to a SC state, inducing 
an AF interface in the process, to reproduce recent experiments \cite{varela}.
Shown are the (a) electronic density $n_{\bf i}$, 
(b) magnetization 
$m_z({\bf i})=n_{{\bf i}\uparrow}-n_{{\bf i}\downarrow}$, and (c) $d$-wave 
superconducting order parameter $\Delta_d({\bf i})$, as a function of layer 
position $i_z$ for two different sublattices [$(-1)^{i_x+i_y}=\pm1$] in the $xy$ 
plane (denoted by circles and crosses). Here ${\bf i}=(i_x,i_y,i_z)$ is a 
lattice site position. The model parameters used are $J_H=8t$, $t_z=t$, 
$W_{\rm L}=10t$, and $n_+^{\rm L}=0.7$ for the left side of the system, and 
$U=4t$, $V=-3t$, $t_z=0.1t$ (to simulate weak hopping between the Cu oxide
layers), $W_{\rm R}=0$, and $n_+^{\rm R}=0.7$ for the 
right side of the system. $\alpha=1$ is set for the whole system which has a size
$L=16\times16\times24$. The interface is located at $i_z=12.5$. 
The localized spins in the left side of the system are fixed to be 
ferromagnetic. The temperature considered here is very low, $T$=$t$/400.
}
\label{fm_af_sc}
\end{center}
\end{figure}

The description of the numerical results starts with a qualitative
reproduction of the recent experiments \cite{varela}, where transfer
of charge from LCMO to YBCO was found to destroy superconductivity at the interface.
In Fig.~\ref{fm_af_sc}, the case when a FM
metal  forms an interface with a superconducting state (as in Fig.~\ref{systems}(a)) is studied. 
The work functions and
parameters are chosen such that a transfer of charge from the left to the right takes
place, as shown in Fig.~\ref{fm_af_sc}(a) where an accumulation of charge at the
interface is found. The band-bending picture, which appears to be true
not only for doped band insulators but also for Hubbard insulators \cite{nagaosa},
suggests that there will be a finite region where the density will be
$n$=1, with the chemical potential moving across the Hubbard gap.
In Fig.~\ref{fm_af_sc}(b) the local magnetization is shown. On the 
left, it is FM as expected in a one-orbital manganite model away from $n$=0 or 1. 
On most of the right, the magnetization is zero as in
a regular superconductor. However, in the interface region the magnetization develops a
staggered character suggesting the presence of AF order. 
Thus, as expected from the mean-field formalism  the density $n=1$ is associated 
with antiferromagnetism. 
Figure~\ref{fm_af_sc}(c) shows the suppression of the superconducting order parameter at the interface.
Overall, these simple results reproduce properly the experiments  where
it was observed that transfer of charge from a manganite LCMO to
YBCO led to the suppression of superconductivity at the interface \cite{varela}.

\subsection{Manganite-Cuprate interface, leading to electron-doped SC}

After having crudely reproduced the main qualitative aspects of previous
experiments, the same formalism can be used  to 
formulate predictions for other systems. Of particular interest is the
combination of an A-type AF state (as it occurs for instance in undoped or highly doped LSMO
and also in bilayer compounds \cite{dagotto-CMR,book})
and an $undoped$ cuprate.
If the transfer of charge occurs in the same directions as before, as
already suggested by the discussion on experimental work functions,
then it would be expected
that at the interface a density larger than $n$=1 would be produced (see the sketch in Fig.~\ref{systems}(b)).
If this doping is as large as 5-10\%, then an electron doped
superconductor could be created in a real system.

\begin{figure}[hbt]
\includegraphics[clip=true,width=7.5cm,angle=-0]{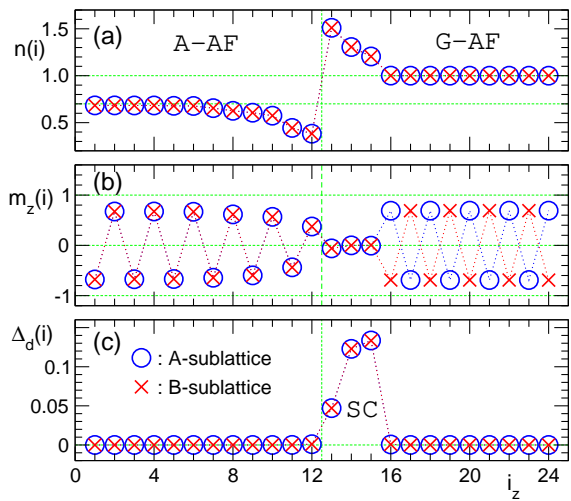}
\begin{center}
\caption{
(Color online) Transfer of charge from an A-type AF state (as it
occurs in some doped manganites \cite{dagotto-CMR})
to an AF insulator (LCO, SCO, NCO, or YBCO), inducing 
an electron-doped SC state at the interface. 
The actual
model parameters 
used here are $J_H=8t$, $t_z=t$, $W_{\rm L}=14t$, and $n_+^{\rm L}=0.7$ for 
the left side of the system, and $U=4t$, $V=-3t$, $t_z=0.1t$, 
$W_{\rm R}=0$, and $n_+^{\rm R}=1.0$ for the right side of the system. 
$\alpha=1$ is set for the whole system, with 
$L=16\times16\times24$ being the lattice studied.
The interface is located at $i_z=12.5$. 
The localized spins in the left side of the systems are fixed to be 
antiferromagnetic in an A-type state, and the temperature of the study was $T$=$t$/400.
}
\label{a-af_sc}
\end{center}
\end{figure}

The actual calculations are conceptually simple and the results are shown in 
Fig.~\ref{a-af_sc}.
In (a), the density is presented: once again
a robust region in parameter space is identified where the transfer of
charge occurs from the AF manganite to the AF cuprate, leading in this case to $n$ larger
than 1 at the interface. In (b), the local magnetization is shown.
In this case, the G-type AF order is seen on most of the right (cuprate) region, but 
this magnetic order disappears at the interface due to the transfer of charge. Finally, in (c)
the superconducting order parameter is shown to become 
nonzero at the interface. Thus,
as anticipated from the introduction, an electron-doped superconductor is predicted to occur
in the manganite/cuprate system described here, within a simple mean-field approximation, for a pair of oxides
with the proper location of chemical potentials.

If the A-type AF manganite state for the left side is replaced by a FM state, as before,
then a similar transfer of charge occurs and within mean-field a SC state is also generated.
However, it is known that the proximity-effect influence of ferromagnetism is detrimental
to singlet superconductivity, thus for experimental realizations of this scenario, a 
non-FM state appears more suitable.

{\it Note}: if the transfer of charge would occur in the other direction, namely
from the $n$=1 AF to a doped FM material, then the cuprate interface would have less
charge and a narrow layer with the properties of a $hole-doped$ cuprate
could be expected. Thus, the simple motive of this effort 
works both ways at the mean-field level.
However, the analysis of work functions from the experimental viewpoint, already discussed,
suggests that in order to generate interfacial superconductivity involving a manganite,
electron doping is the most likely outcome since the chemical potential of LSMO is
above that of LCO, NCO, SCO, and YBCO.

\subsection{SC at BI-AF and Metal-AF interfaces}

\begin{figure}[hbt]
\includegraphics[clip=true,width=7.5cm,angle=-0]{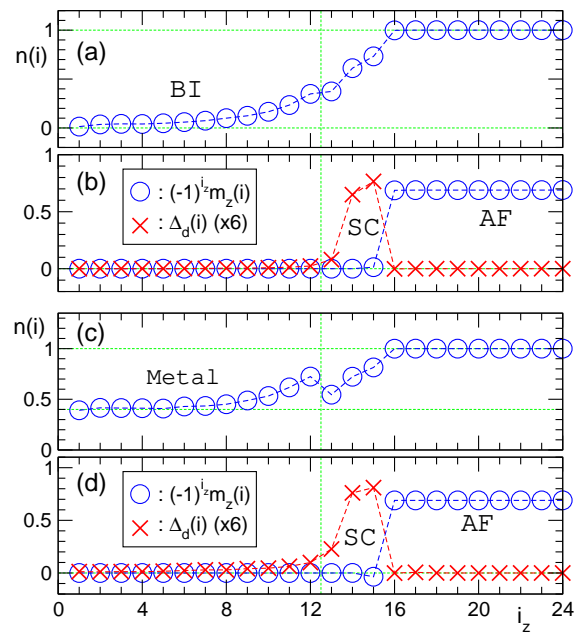}
\begin{center}
\caption{
(Color online) Generation of a SC state at the interface between a band 
insulator (BI) and an AF insulator and also between a standard metal and an AF. 
As in other figures, shown are the
electron density $n_{\bf i}$ 
[(a) and (c)], and staggered 
magnetization $(-1)^{i_z}m_z({\bf i})$ and $d$-wave superconducting order 
parameter $\Delta_d({\bf i})$ [(b) and (d)] as a function of layer position 
$i_z$. 
The model parameters used in (a) and (b) are $J_H=0$ (to avoid ferromagnetism), $t_z=t$, 
$W_{\rm L}=0$, and $n_+^{\rm L}=0.0$ (to get a crude band insulator) 
for the left side of the system, and 
$U=4t$, $V=-3t$, $t_z=0.1t$, $W_{\rm R}=0$, and $n_+^{\rm R}=1.0$ 
for the right 
side of the system. The same parameters are used in (c) and (d) except 
$n_+^{\rm L}=0.4$ for the left side of the system. 
$\alpha=1$ is set for the whole system with 
$L=16\times16\times24$ being the lattice studied, 
and the interface is located at $i_z=12.5$.
}
\label{bi_af}
\end{center}
\end{figure}

The generation of a superconductor via transfer of charge from another
compound is certainly not restricted to occur only when LSMO is involved, but 
it should happen under far more general circumstances. For instance, 
the case of a band-insulator (BI) forming an interface with an AF
was also studied.
Figure~\ref{bi_af}(a) shows the density profile, indicating that the parameters of
the calculation are such that the
transfer of charge this time occurs from the AF to the BI, inducing
a region in the cuprate with hole doping, and concomitant superconductivity
(as shown in (b)). Certainly, electron-doped SC can be induced as well by
adjusting the work functions.

If instead of a BI, a standard metal is used
(modeled in our study by merely using Hubbard $U$=0 in a tight-binding model),
then for appropriate work functions also a charge transfer is to be expected
(see (c) and (d)). Thus, the most important aspect of the problem is
to identify materials with the proper relative location of work functions such that the
charge transfer occurs in the proper direction, and also such that there
is a good matching between lattice spacings to avoid generating extra
complications for the transfer of charge to occur.

\section{Extended Dynamical Mean Field Theory Results for Hubbard Model Interfaces}

To place our findings on a firmer footing, we describe here
the results obtained from cluster dynamical mean field theory
(CDMFT) \cite{kotliar2001} generalized to a layered geometry for the 
case of a ferromagnetic -- Mott insulator
interface. The previous section showed that there are many similarities
in the use of FM or AF manganites in the heterostructure, since their main role is the donation of
carriers to the cuprate. Since technically the case of FM is simpler, in this
section we focus on a FM manganite.
CDMFT is a powerful technique which maps a full many-body
problem onto that of a cluster embedded in a self-consistent
medium. The model Hamiltonian considered here is identical to 
Eqs.~1 and 2, with the addition of the local Coulomb interaction represented by the
Hubbard $U$ on the Mott insulating side of the interface (and without the
explicitly attractive nearest-neighbors attraction) . Long-range
Coulomb interactions are treated again using the Poisson equation as
in Eq.~4.

CDMFT can be generalized to treat the effects of a layered geometry in
the same manner as DMFT was adapted to treat a film
geometry \cite{potthoff2003,okamoto2004}. Each layer is mapped onto an
independent cluster impurity embedded in a medium. The solution for
the self-energy of each layer is then used via a self-consistency
condition to obtain the local Green's function of the entire
lattice. Hopping between the layers is taken into account only at the
level of the self-consistency condition. To make the problem
computationally feasible, we fix the number of layers to $L_{\rm
FM}=5$ and $L_{\rm MI}=5$ for the ferromagnet and the Mott insulator,
respectively.  Further, each layer is mapped to a 4-site cluster
embedded in a bath of 8 sites. SC correlations are included in the
bath around the MI layers to allow for the SC instability and a paramagnetic
solution is imposed. Individual
layers are oriented in the $xy$-plane and stacked in the
$z$-direction. The algorithm starts by making an initial guess for the
local chemical potential and the bath parameters across the layered
structure. Solving the cluster impurity model in each layer using the
Lanczos method we obtain the cluster Green's function, cluster density,
and self-energy. The density in each layer is used to obtain the new
local potential $\phi_z$ using the Poisson equation (Eq.~4). The
self-energy in each layer is used to obtain a local Green's function
for the lattice using the equation below:
\begin{equation}
G_{\rm loc}(z,\omega)=\int\frac{dK_xdK_y}{\pi^2}G(z,z',K_x,K_y,\omega)
\end{equation}
where
\begin{equation}
\begin{array}{l}
\displaystyle{
G(z,z',K_x,K_y,\omega)= 
} \\
\displaystyle{
[\omega+\mu_z-t(K_x,K_y,z,z')-\Sigma(z,z',K_x,K_y,\omega)]^{-1}.
}
\end{array}
\end{equation}
Here, $t(K_x,K_y,z,z')$ denotes the Fourier transform of the hopping
matrix (both intralayer and interlayer), $K_x$ and $K_y$ denote the
superlattice momenta within each layer, and $z$ denotes the layer
index. $\mu_z$ includes all the terms that couple to the density,
including the chemical potential, $\phi_z$, site potential,
and the work function on either side of the interface. 
The Green's function $G_{\rm loc}(z,\omega)$ is
combined with the Dyson equation to give a new Weiss field that is
then used to obtain a new set of bath parameters using a conjugate
gradient minimization. Convergence is achieved when the bath
parameters for each layer and the density profile across the layers do
not change with further iterations.
\begin{figure}[ht]
\begin{center}
\includegraphics[width=7.0cm,angle=-0] {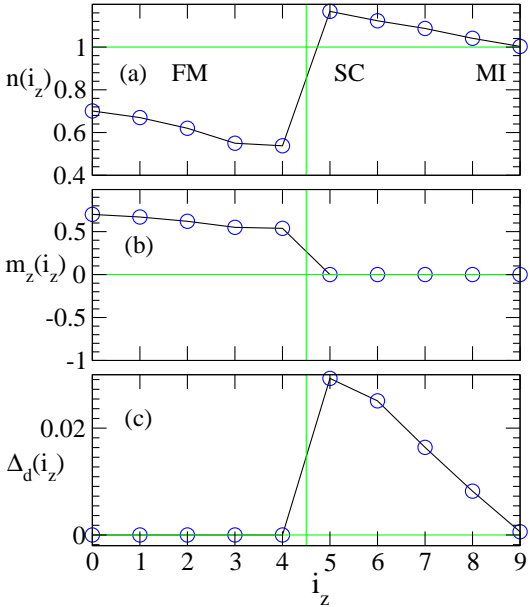}
\end{center}
\caption{Transfer of charge from a FM manganite to a Mott insulating
cuprate inducing a $d$-wave SC
state at the interface in a 10-layer heterostructure, obtained using CDMFT
generalized to a layered geometry. (a) is the electronic density profile
$n_z$; (b)  is the magnetic moment in each layer $z$, denote by $m_z(i_z)$; 
and (c) is the $d$-wave
SC order parameter, $\Delta_d(i_z)=\langle c_{z,i\uparrow}c_{z,i+x\downarrow}\rangle$
across the heterostructure. The model parameters used are $J_H=8t$, $t_z=t$, $W_L=12t$,
$n_{+}^{L}=0.7$, $n_{+}^{R}=1.0$, $W_R=0$, $\alpha=1$, and $U/t$=8.}
\label{interface}
\end{figure}

In Fig.~\ref{interface}(a) the density profile is presented. The bulk
density is at $n$=0.7 on the left side (Mn oxide)
and $n$=1.0 on the right (Cu oxide, $U/t$=8). As expected from
the previous discussions, electron doping of the Mott insulator 
occurs at the interface due to a transfer of charge
from the manganite to the cuprate. 
Figure~\ref{interface}(b) provides the magnetization at each layer. Note
that the effects of penetration of the FM moment onto the Mott insulator side are not considered
in this calculation. In a real setup, it is likely that a finite polarization will be found
on the right hand side of the figure.
Figure~\ref{interface}(c)
shows the $d$-wave SC order parameter, which  in this heterostructure 
becomes  non-zero in the cuprate's
interfacial region and decays rapidly into the bulk of the Mott insulator. Thus, an
electron-doped $d$-wave superconductor can be realized at the interface
of a manganite and a cuprate, due to the expected transfer of charge
between them, even when dynamical effects in the mean-field approximation
are taken into account.

\section{Discussion and Summary}  

In this manuscript, the possible charge transfer from a manganite to an undoped cuprate was
discussed in the context of oxide heterostructures. This issue is nontrivial since $a$ $priori$
the existence of a robust gap in the undoped 
Cu oxides would have suggested the lack of available states for manganite electrons to pour
into the antiferromagnetic cuprates near the interfaces. The recently discovered $indirect$ nature
of the Cu-oxides gaps effectively reduces the magnitude of the Hubbard gap. Taking this into
consideration, our analysis suggests that the Fermi level of manganites, in a robust range of hole
doping, lies above the chemical potential of several cuprates and even above the bottom of the upper
Hubbard band of SCO and NCO.
This opens the possibility
of electron doping of high-$T_c$'s at interfaces with manganites (and other oxides).
Under idealized conditions, the doping discussed here may induce a superconducting state (if the
manganite is not ferromagnetic), an exotic metallic state containing polarized carriers in an
antiferromagnetic background (if the manganite is ferromagnetic), 
or still a superconducting state,
likely with a triplet component, if the spin polarization of the carriers coming from the manganite
is only partial.

An important component of our effort is the introduction of a systematic procedure to analyze 
photoemission and diffusion voltage experiments to predict the direction of flow of charge at
interfaces. A figure with the relative Fermi levels and gap locations of several oxides was presented.

\begin{figure}[hbt]
\includegraphics[clip=true,width=7.5cm,angle=-0]{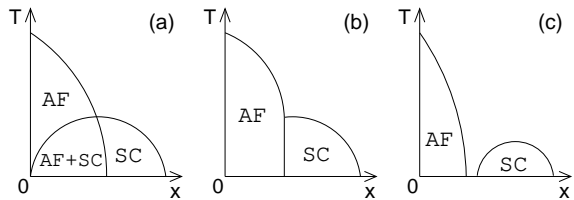}
\begin{center}
\caption{ (a) and (b) are the possible phase diagrams of the cuprates in the clean limit \cite{alvarez},
which may be experimentally realized at the interfaces discussed in this paper. No distinction is
made between hole or electron doping, $x$ represents both. (c) is the well-known phase diagram of chemically
doped LSCO. According to Ref.~\cite{alvarez}, the glassy state between AF and SC phases is caused by
quenched disorder.
}
\label{phase_SC}
\end{center}
\end{figure}

If the superconducting state is ever realized in clean interfaces, a fundamental issue to address
is the unveiling of the phase diagram of cuprates in the absence of quenched disorder. This is true for
both electron and hole doping. Recent phenomenological calculations \cite{alvarez} involving noninteracting carriers
in interaction with the AF and SC order parameters (an extension of the traditional Landau-Ginzburg approach)
suggest that these phases should be separated by either a region of local coexistence or a first-order
transition. A glassy state, as in the widely studied phase diagram of LSCO, was reproduced $only$ incorporating
quenched disorder. For a sketch of these diagrams see Fig.~\ref{phase_SC}. The generation of superconductivity
at interfaces may reveal the true phase diagram of clean cuprates.

It is clear that our calculations can only be considered  as suggestive, and its main purpose is to induce further
work in this area. Important simplifications employed in the study,
including the neglect of lattice reconstructions, vacancies, and polarity effects, need to be addressed.
$Ab$ $initio$ calculations are crucial to clarify these issues. And, of course, the experimental
realization of the interfaces discussed here would provide a definitive answer to the proposal
of electron doping of cuprates at the interfaces.

\section{Acknowledgment}
We thank I. Bozovic, H. Y. Hwang, M. Kawasaki, T. Kopp,
H. Kumigashira, S. Pennycook, A. Sawa, Y. Tokura, 
and M. Varela for valuable discussions. This work was supported 
in part by the NSF grant DMR-0443144 and by the Division of Materials Sciences and
Engineering, Office of Basic Energy Sciences, U.S. Department of Energy,
under contract DE-AC05-00OR22725 with Oak Ridge National Laboratory,
managed and operated by UT-Battelle, LLC.
Support by the LDRD program at ORNL is also acknowledged.



\begin{thebibliography}{99}
\bibitem{ohtomo02} A. Ohtomo, D. A. Muller, J. L. Grazul, and H. Y. Hwang, Nature {\bf 419}, 378 (2002).

\bibitem{ohtomo04}A. Ohtomo and H. Y. Hwang, Nature {\bf 427}, 423 (2004).

\bibitem{okamoto04a} S. Okamoto and A. Millis, Nature {\bf 428}, 630 (2004).

\bibitem{okamoto04b} S. Okamoto and A. J. Millis, Phys. Rev. B {\bf 70}, 075101 (2004).

\bibitem{okamoto2004} S. Okamoto and A. J. Millis, Phys. Rev. B
{\bf 70}, 241104(R) (2004).

\bibitem{nakagawa}N. Nakagawa, H. Y. Hwang, and D. A. Muller, Nature Materials {\bf 5}, 204 (2006),
and references therein.

\bibitem{thiel} S. Thiel, G. Hammerl, A. Schmehl, C. W. Schneider, and J. Mannhart, Science {\bf 313}, 1942 (2006),
and references therein.

\bibitem{dagotto} E. Dagotto, Science {\bf 309}, 257 (2005), and references therein.


\bibitem{yoshida} A. F. Volkov, F. S. Bergeret and K. B. Efetov, 
Phys. Rev. Lett. {\bf 90}, 117006 (2003). See also
N. Yoshida, M. Fogelstrom, cond-mat/0511009.

\bibitem{eschrig} M. Eschrig, J. Kopu, J. C. Cuevas, and Gerd Schon,  Phys. Rev. Lett. {\bf 90}, 137003 (2003).

\bibitem{kontos} T. Kontos, M. Aprili, J. Lesueur, F. Genet, B. Stephanidis, and R. Boursier, Phys. Rev. Lett. {\bf 89}, 137007 (2002).

\bibitem{vasko}
V. A. Vas'ko, V. A. Larkin, P. A. Kraus, K. R. Nikolaev, D. E. Grupp, C. A. Nordman, and A. M. Goldman, 
Phys. Rev. Lett. 78, 1134 (1997).

\bibitem{melo} C. A. R. Sa de Melo, Phys. Rev. Lett. {\bf 79}, 1933 (1997).

\bibitem{dagotto-CMR} E. Dagotto, T. Hotta, and A. Moreo, Physics Reports {\bf 344}, 1 (2001). 

\bibitem{book} E. Dagotto, {\it Nanoscale Phase Separation and Colossal Magnetoresistance}, Springer (2002).


\bibitem{mannhart} J. Mannhart, D. G. Schlom, J. G. Bednorz, and 
K. A. Muller, Phys. Rev. Lett. {\bf 67}, 2099 (1991).

\bibitem{ahn} C. H. Ahn, S. Gariglio, P. Paruch, T. Tybell, 
L. Antognazza, and J.-M. Triscone, Science {\bf 284}, 1152 (1999).

\bibitem{pavlenko} For recent work in this area see
N. Pavlenko, I. Elfimov, T. Kopp, and G. A. Sawatzky, cond-mat/0605589, and references therein.
For related literature see: V. Koerting, Q. Yuan, P.J. Hirschfeld, T. Kopp, and J. Mannhart,
Phys. Rev. B {\bf 71}, 104510 (2005); N. Pavlenko and T. Kopp,
Phys. Rev. B {\bf 72}, 174516 (2005).


\bibitem{varela} M. Varela, A.R. Lupini, V. Pe\~na, Z. Sefrioui, I. Arslan, 
N.D. Browning, J. Santamaria, S.J. Pennycook, cond-mat/0508564; see also
Todd Holden, H.-U. Habermeier, G. Cristiani, A. Golnik, A. Boris, A. Pimenov, 
J. Humlicek, O. I. Lebedev, G. Van Tendeloo, B. Keimer, and C. Bernhard, 
Phys. Rev. B {\bf 69}, 064505 (2004);
A. Hoffmann, S. G. E. te Velthuis, Z. Sefrioui, J. Santamaria, 
M. R. Fitzsimmons, S. Park, and M. Varela, 
Phys. Rev. B{\bf 72}, 140407(R) (2005);
V. Pe\~na, C. Visani, J. Garcia-Barriocanal, D. Arias, Z. Sefrioui, 
C. Leon, J. Santamaria, and C. A. Almasan, 
Phys. Rev. B {\bf 73}, 104513 (2006).

\bibitem{hilgenkamp} H. Hilgenkamp and J. Mannhart, Rev. Mod. Phys. {\bf 74}, 485 (2002); and references therein.
See also B. Nikolic, J. K. Freericks, and P. Miller, Phys. Rev. B {\bf 65}, 064529 (2002); U. Schwingenschlogl and
C. Schuster, cond-mat/0702098.

\bibitem{sefrioui} Z. Sefrioui, D. Arias, V. Pe\~na, J. E. Villegas, M. Varela, P. Prieto, C. Le\'on, J. L. Martinez,
and J. Santamaria, Phys. Rev. B {\bf 67}, 214511 (2003). See also V. Pe\~na, Z. Sefrioui, D. Arias, C. Le\'on, J. Santamaria,
J. L. Martinez, S. G. E. te Velthuis, and A. Hoffmann, Phys. Rev. Lett. {\bf 94}, 057002 (2005); J. Chakhalian, J. W. Freeland,
G. Srajer, J. Strempfer, G. Khaliullin, J. C. Cezar, T. Charlton, R. Dalgliesh, C. Bernhard, G. Cristiani, H-U. Habermeier, and
B. Keimer, Nature Physics {\bf 2}, 244 (2006).

\bibitem{oh} The $hole$ doping of an undoped cuprate when in combination with STO has already been reported in
S. Oh, M. Warusawithana, and J. N. Eckstein, Phys. Rev. B {\bf 70}, 064509 (2004). See also
J. N. Eckstein and I. Bozovic, Annu. Rev. Mater. Sci. {\bf 25}, 679 (1995); G. Yu. Logvenov, A. Sawa, C. W. Schneider and
J. Mannhart, App. Phys. lett. {\bf 83}, 3528 (2003); and references therein. 

\bibitem{brey} L. Brey, cond-mat/0611594, and references therein.


\bibitem{Schulte01}K. Schulte, M. A. James, L. H. Tjeng, P. G. Steeneken, G. A. Sawatzky, R. Suryanarayanan, G. Dhalenne, and A. Revcolevschi, 
Phys. Rev. B {\bf 64}, 134428 (2001).  

\bibitem{Suzuki89}M. Suzuki, Phys. Rev. B {\bf 39}, 2312 (1989). 

\bibitem{Ino00} A. Ino, C. Kim, M. Nakamura, T. Yoshida, T. Mizokawa, Z-X. Shen, A. Fujimori, T. Kakeshita, H. Eisaki, 
and S. Uchida, Phys. Rev. B {\bf 62}, 4137 (2000).

\bibitem{polarons}O. R{\"o}sch, O. Gunnarsson, X. J. Zhou, T. Yoshida, T. Sasagawa, A. Fujimori, 
Z. Hussain, Z.-X. Shen, and S. Uchida, Phys. Rev. Lett. {\bf 95}, 227002 (2005).

\bibitem{Allen90}J. W. Allen, C. G. Olson, M. B. Maple, J.-S. Kang, L. Z. Liu, J.-H. Park, R. O. Anderson, W. P. Ellis, J. T. Markert, 
Y. Dalichaouch, and R. Liu, Phys. Rev. Lett. {\bf 64}, 595 (1990). 

\bibitem{Namatame90}H. Namatame, A. Fujimori, Y. Tokura, M. Nakamura, K. Yamaguchi, A. Misu, H. Matsubara, S. Suga, H. Eisaki, 
T. Ito, H. Takagi, and S. Uchida, Phys. Rev. B {\bf 41}, 7205 (1990). 

\bibitem{alternative} Note that the chemical potential difference between
LCO and NCO is still controversial. Other estimations 
arrive to a value $\sim 1$ eV
for this quantity. See P. G. Steeneken, L. H. Tjeng, G. A. Sawatzky, A. Tanaka, 
O. Tjernberg, G. Ghiringhelli, N. B. Brookes, A. A. Nugroho, and A. A. Menovsky,
Phys. Rev. Lett. {\bf 90}, 247005 (2003).

\bibitem{Fujimori02}A. Fujimori, A. Ino, J. Matsuno, T. Yoshida, K. Tanaka, and T. Mizokawa,         
J. Electr. Spectrosc. Relat. Phenom. {\bf 124}, 127 (2002). 

\bibitem{Tajima89}S. Tajima, H. Ishii, T. Nakamura,H. Takagi, S. Uchida, M. Seki, S. Suga, Y. Hidaka, M. Suzuki, T. Murakami, 
K. Oka, and H. Unoki, J. Opt. Soc. Am. B {\bf 6}, 475 (1989). 

\bibitem{Tsutsui99}K. Tsutsui, T. Tohyama, and S. Maekawa, Phys. Rev. Lett. {\bf 83}, 3705 (1999). 

\bibitem{Hasan00}M. Z. Hasan, E. D. Isaacs, Z.-X. Shen, L. L. Miller, K. Tsutsui, T. Tohyama, and S. Maekawa, Science {\bf 288}, 1811 (2000).

\bibitem{BIS98} A. Fujimori, A. Ino, T. Mizokawa, C. Kim, Z.-X. Shen, 
T. Sasagawa, T. Kimura, K. Kishio, M. Takaba, K. Tamasaku, H. Eisaki 
and S. Uchida, J. Phys. Chem. Solids {\bf 59}, 1892 (1998).

\bibitem{Muraoka04}Y. Muraoka, T. Muramatsu, J. Yamaura, and Z. Hiroi, Appl. Phys. Lett. {\bf 85}, 2950 (2004). 

\bibitem{Nakamura07}M. Nakamura, A. Sawa, H. Sato, H. Akoh, M. Kawasaki, and Y. Tokura, Phys. Rev. B {\bf 75}, 155103 (2007). 

\bibitem{SawaPC}T. Fujii, M. Kawasaki, A. Sawa, Y. Kawazoe, H. Akoh, and Y. Tokura, Phys. Rev. B (2007), in press. 

\bibitem{Muramatsu05}T. Muramatsu, Y. Muraoka, and Z. Hiroi, Jpn. J. Appl. Phys. {\bf 44}, 7367 (2005). 

\bibitem{Matsuno02}J. Matsuno, A. Fujimori, Y. Takeda and M. Takano, Europhys. Lett. {\bf 59}, 252 (2002). 

\bibitem{Ino97}A. Ino, T. Mizokawa, A. Fujimori, K. Tamasaku, H. Eisaki, S. Uchida, T. Kimura, T. Sasagawa, and K. Kishio, 
Phys. Rev. Lett. {\bf 79}, 2101 (1997). 

\bibitem{Harima03}N. Harima, A. Fujimori, T. Sugaya, and I. Terasaki, Phys. Rev. B {\bf 67}, 172501 (2003). 

\bibitem{Yagi06}H. Yagi, T. Yoshida, A. Fujimori, Y. Kohsaka, M. Misawa, T. Sasagawa, H. Takagi, M. Azuma and M. Takano, Phys. Rev. B {\bf 73}, 172503 (2006). 

\bibitem{Harima01}N. Harima, J. Matsuno, A. Fujimori, Y. Onose, Y. Taguchi, and Y. Tokura, Phys. Rev. B {\bf 64}, 220507(R) (2001). 

\bibitem{Jong03}M. P. de Jong, V. A. Dediu, C. Taliani, and W. R. Salaneck, J. Appl. Phys. {\bf 94}, 7292 (2003). 

\bibitem{Minohara07}M. Minohara, I. Ohkuho, H. Kumigashira, and M. Oshima, 
Appl. Phys. Lett. {\bf 90}, 132123 (2007).

\bibitem{Reagor04}W. Reagor, S. Y. Lee, Y. Li, and Q. X. Jia, J. Appl. Phys. {\bf 95}, 7971 (2004). 

\bibitem{YBCO-work} The work function of YBCO has been experimentally measured. According to T. Hirano,
M. Ueda, K. Matsui, T. Fujii, K. Sakuta, and T. Kobayashi, Jpn. J. Appl. Phys. {\bf 31}, L1345 (1992),
the YBCO work function is 6.1 eV. Since the work function of LSMO 
is $\sim 4.8$~eV \cite{Jong03,Minohara07,Reagor04}, their difference is 1.3 eV, which is not too different (given
the uncertainties in these experimental results) from the
numbers deduced from Fig.~\ref{fig:LCO_SCO_NbSTO_YBCO_LSMO}.


\bibitem{STO-comment}
Are there other transition-metal oxides that can replace manganites in this context?
Consider first the case of ruthenates. In addition to the $3d$ oxides, clearly the
$4d$ transition-metal oxides are another potentially important class of materials for our purposes. 
Among them, SrRuO$_3$ (SRO) is known to be highly conductive and, therefore, a potential candidate 
for electrode in transition-metal based devices.
However, 
through permittivity measurements, the work function of (001) SRO is estimated to be 5.2~eV (see
X. Fang and T. Kobayashi, Appl. Phys. A {\bf 69}, S587 (1999)).
This SRO work function is rather large compared with the value 
$\sim 3.5$~eV of the double-layered La$_{1.2}$Sr$_{1.8}$Mn$_2$O$_7$ 
and $\sim4.8$ of the cubic perovskite LSMO \cite{Minohara07}. 
This may indicate the following trend: the larger the atomic number of the transition-metal is, 
the larger the work function becomes. 
In other words, if one wants to inject electrons to some material, one may need to 
use transition-metal oxides with $smaller$ atomic number than manganites (since reducing the
atomic number also reduces the electro-negativity)
From this point of view, electron-rich titanates (via the addition of Nb) 
can provide another route for electron doping of cuprate parent compounds, 
as very recently experimentally reported \cite{Nakamura07}.
Our study is by no means restricted to manganites as electron
donors.


\bibitem{tuv} E. Dagotto, J. Riera, Y. C. Chen, A. Moreo, A. Nazarenko, 
F. Alcaraz, and F. Ortolani, \prb {\bf 49}, 3548 (1994); and
references therein. See also  R. Micnas, J. Ranninger, and S. Robaszkiewicz, Rev. Mod. Phys. {\bf 62}, 
113 (1990); I. Martin, G. Ortiz, A. V. Balatsky, and A. R. Bishop, Int. J. of Mod. Phys. {\bf 14}, 3567 (2000).

\bibitem{alvarez} G. Alvarez, M. Mayr, A. Moreo, and E. Dagotto, Phys. Rev. B {\bf 71}, 014514 (2005).  See also
M. Mayr, G. Alvarez, A. Moreo, and E. Dagotto, Phys. Rev. B {\bf 73}, 014509 (2006). 


\bibitem{datta} See, for example, S. Datta, 
{\it Electronic transport in mesoscopic systems} 
(Cambridge University Press, Cambridge, 1997).

\bibitem{bdg} See, e.g., P. G. De Gennes 
{\it Superconductivity of Metals and Alloys} 
(Perseus Books, New York, 1999). 


\bibitem{nagaosa}  T. Oka and N. Nagaosa,
Phys. Rev. Lett. {\bf 95}, 266403 (2005).


\bibitem{kotliar2001} G. Kotliar, S. Y. Savrasov, G. Palsson, and G. Biroli
Phys. Rev. Lett. {\bf 87}, 186401 (2001).

\bibitem{potthoff2003} M. Potthoff and W. Nolting, Phys. Rev. B {\bf 59}, 2549; 
{\bf 60}, 7834 (1999);
S. Schwieger, M. Potthoff, and W. Nolting,
Phys. Rev. B {\bf 67}, 165408 (2003).
































\end{thebibliography}
\end{document}